\documentclass[aps,prc,showpacs,twocolumn,preprintnumbers,floatfix,
showkeys,tightenlines]{revtex4}
\usepackage{amsmath,amssymb,amsfonts}
\usepackage{graphicx}
\usepackage{color}

\begin{document}
\title{ Colour-singlet clustering of partons and recombination model
for hadronization of quark-gluon plasma}
\author{Raktim \surname{Abir}} 
\email{raktim.abir@saha.ac.in}
\author{Munshi G. \surname{Mustafa}} 
\email{munshigolam.mustafa@saha.ac.in}
\affiliation{
Theory Division, Saha Institute of Nuclear Physics, 1/AF Bidhannagar, Kolkata 700064, India\\
}

\begin{abstract}
$SU(N_c)$ colour-singlet restriction, along with flavour and spin symmetry, 
on thermal partonic ensemble is shown to recombine the partons with internal
colour structure into  colour-singlet multi-quark clusters which
can be identified with various hadronic modes at a given temperature.
This can provide a possible basis for recombination model for hadronization of 
quark-gluon plasma. This also leads to 
a natural explanation for the ratio of (anti)protons to pions and the
quark number scaling of the elliptic flow coefficient in relativistic 
heavy-ion collisions. 
  
\end{abstract}
%\preprint{SINP/TNP/06-30}
\maketitle

\noindent{\it Introduction:}
By now it is generally believed that the deconfinement of strongly interacting
matter has been  achieved in relativistic heavy-ion collisions~\cite{white}.
Remarkable confirmations for this premise come from elliptic flow of hadrons
and its scaling with the number of valence quarks~\cite{ellip}, jet 
quenching~\cite{jet}, and radiation of thermal photons~\cite{phot}. However, 
the results from RHIC experiments also revealed some interesting facts that 
the nuclear suppression factor depends on the hadron species~\cite{jet} and 
the proton to pion ratio~\cite{phenix} has a plateau around unity in 
the transverse momentum range (2 - 4) GeV/$c$.  Recently, it has been 
argued in Refs.~\cite{bass,greco} that the hadron production at low 
momenta in a dense medium takes place through the recombination of 
partons which explains some of the surprising results of RHIC experiments.

Generally, hadronization by recombination is a $``$two to one" process 
in a medium where the 
(anti)quarks are effective degrees of freedom and gluons are dynamical 
ones that disappear at hadronization. This simply boils down to the fact 
of {\it a correct counting of quantum states and momenta} within a 
dynamical theory. However, it is an extremely difficult task within the 
dynamical QCD and various models~\cite{rec,rec1,bass,greco,cmko} have been 
formulated to describe the hadron productions in heavy-ion collisions.

Recently, under a sudden approximation the recombination is considered with a
perturbative quark, {\it i.e.}, minijet and a thermal (anti)quark~\cite{greco} 
whereas in Ref.~\cite{bass} it is with thermal (anti)quarks. 
In Ref.~\cite{greco} it is argued that an additional contribution is 
necessary for the transverse momentum spectra of hadron at the transition 
region between the thermal recombination and the individual fragmentation. 
Such contribution comes from the recombination of a minijet and a thermal 
quark. In Ref.~\cite{bass}, on the
other hand, it is argued that for momenta below 5 GeV/$c$ the thermal
recombination dominates whereas beyond $5$ GeV/$c$ the fragmentation of 
independent minijet dominates the hadron production. Also, the competition 
between the recombination and fragmentation pushes the onset of 
fragmentation to relatively higher transverse momentum of $5-6$ GeV/$c$.
This indicates that the quark recombination phenomena for hadronisation of
quark-gluon plasma remain an open as well an interesting problem.

In the present article, we show that the recombination 
phenomena arise spontaneously upon application of colour-singlet 
projection operator on the partition function for an assembly of quarks 
and antiquarks having internal 
colour structures. Such recombination phenomena naturally explain the 
baryon-to-meson ratio and the azimuthal anisotropy of hadron distributions 
that scales with the number of valence quarks. 

\noindent {\it{Quantum statistical mechanics and colour-singlet ensemble:}}
The statistical behaviour of a quantum gas in thermal equilibrium
is usually studied through an appropriate ensemble. In general one defines
a density matrix for the system as
\begin{equation}
{\rho(\beta)} =  {\rm {exp}} (-\beta
{\hat H}) \ \ , \label{denmat}
\end{equation}
where $\beta=1/T$ is the inverse of temperature and ${\hat H}$ is the
Hamiltonian of the physical system. The corresponding partition function 
for a quantum gas having a finite volume can be written as
\begin{equation}
{\cal Z} = {\rm{Tr}} \left ( {\hat {\cal P}} e^{-\beta 
{\hat H}} \right ) = \sum_n \left \langle n \left | 
{\hat {\cal P}} e^{-\beta {\hat H}} 
\right | n \right \rangle \ \ , \label{part} 
\end{equation}
where $|n\rangle $ is a many-particle state in the Hilbert space 
${\cal H}$ and 
${\hat {\cal P}}$ is the projection operator for any desired
configuration. We propose to consider the statistical 
thermodynamical description of a quantum gas consisting of quarks and 
antiquarks, such that the underlying symmetry amounts to 
reordering of the partition function in terms of {\it  the  
colour-singlet multi-quark modes} at a given temperature. 
We also assume here that the  gluons are in thermal background 
and the partonic matter is mainly composed of quarks and antiquarks.

Now for a symmetry group ${\cal G}$ (compact Lie group) 
having unitary representation ${\hat U}(g)$ in a Hilbert space
${\cal H}$, the projection operator can be written as \cite{weyl,aub}
\begin{equation}
{\hat{\cal P}}_j = d_j \int_{\cal G} {\rm d}\mu (g) \chi^{\star}_j(g)
{\hat U}(g) \ \ , \label{proj}
\end{equation}
where $d_j$ and $\chi_j$ are, respectively, the dimension and the
character of the irreducible representation $j$ of ${\cal G}$. 
${\rm d}\mu (g)$ is the normalised Haar measure in the group
${\cal G}$. The
symmetry group associated with the colour-singlet configuration of
the system is $SU(N_c)$, $N_c$ is the number of colour corresponding to
fundamental representation.  For the $SU(N_c)$ colour-singlet configuration
$d_j=1$ and $\chi_j=1$. 
Now the colour-singlet partition function for the system becomes,
\begin{equation}
{\cal Z}_S = {\rm{Tr}} \left ( \int_{SU(N_c)} {\rm d}\mu (g)\ \  
{\hat U}(g)\ {\rm {exp}} (-\beta {\hat H}) \right ) \ \ , \label{part0} 
\end{equation}
where ${\hat U}(g)$ can be thought of a local link variable that links
the (anti)quarks in a given state of the physical system. Now,  
the trace in a Fock space results in a product of the fermionic 
determinant over momentum modes and colour degrees of freedom, which we
shall see later.  Thus, the interchange of the $`${\rm {Tr}}' and the 
integration will lead to an overall product over momentum  
and color states and one can write the (\ref{part0}) as 
\begin{equation}
{\cal Z}_S =\prod \int_{SU(N_c)} {\rm d}\mu (g) {\rm{Tr}} 
\left ( {\hat U}(g) 
{\rm {exp}} (-\beta {\hat H}) \right ) \ \ . \label{part1} 
\end{equation}

We neglect the mutual interactions among the constituents,
although they must interact in order to come to a thermal equilibrium.
One can imagine a situation by first allowing them to come to a thermal 
equilibrium and then slowly turning off the interactions~\cite{Gale}. 
Such a simple thermodynamic description is often useful for various 
physical systems ({\it e.g.}, electrons in metal, blackbody photons 
in a heated cavity, phonons at low temperature, neutron matter 
in neutron stars, etc.). The full Hamiltonian is 
then the sum of the Hamiltonians for each species, {\it i.e.}, quarks
and antiquarks as ${\hat H}={\hat H}_q+{\hat H}_{\bar q}$ with 
${\hat H}_i= {\hat h}_i-\mu_i{\hat N}_i$, in the grand canonical 
ensemble with the usual meaning of $\mu_i$ and ${\hat N}_i$.

Now, the Hilbert space ${\cal H}$ of the composite system has a structure 
of a tensor product of the individual Fock spaces of quarks and antiquarks 
as
${\cal H} = {\cal H}_q \otimes {\cal H}_{\bar q}$,
where the subscripts $q$ and $\bar q$ denote the quark and antiquark, 
respectively.
Due to this the partition function in (\ref{part1}) in Hilbert
space ${\cal H}$ decomposes into the product of two 
traces~\cite{aub,red,mus0,mus} in their respective Fock spaces as
\begin{eqnarray}
{\cal Z}_S{\hspace*{-0.09in}} &=&{\hspace*{-0.10in}} \prod
\int_{SU(N_c)} {\hspace*{-0.32in}} 
 {\rm d}\mu (g) \ {\rm{Tr}} \left ( {\hat U}_q(g) 
 e^{-\beta {\hat H}_q} \right )   
 {\rm{Tr}} \left ( {\hat U}_{\bar q}(g) 
 e^{-\beta {\hat H}_{\bar q}} \right ) .  \label{part2}
\end{eqnarray}
For simplicity, we approximate the local link variable $U(g)$ to be 
diagonal matrix related to {\it only} the diagonal generators in colour 
space associated with maximal abelian sub-group (Cartan space)~\cite{weyl} 
of $SU(N_c)$. As for example, $SU(3)$ colour gauge group has only two 
parameter abelian sub-group associated with the two diagonal generators 
that would characterise $U(g)$, including its diagonalisation as can be seen
below. Under this approximation the Haar measure corresponding to $SU(N_c)$ 
can now be written in the Weyl reduced~\cite{weyl,aub} form as
\begin{eqnarray}
\int_{SU(N_c)} {\rm d}\mu (g) &=& 
 \frac{1}{N_c!}\left ( \prod^{N_c}_{l=1} 
\int^\pi_{-\pi} \frac{{\rm d}\theta_l}{2\pi}\right )
\delta\left(\sum_l^{N_c}\theta_l\right)  \nonumber \\ 
&& \times J(e^{i\theta_1}\cdot\cdot e^{i\theta_{N_c}}) , 
\label{measure}
\end{eqnarray}
where $J$ is the Jacobian of transformation (also known as Vandermonde 
determinant~\cite{jacob}). $\theta_l$ is a class parameter characterizing 
the group element $g$ such that $U(g)$ can be diagonalised. It also obeys 
the periodicity condition $\sum^{N_c}_{l=1}\theta_l=0 \ ({\rm{mod}} 2\pi)$, 
which ensures that the group element is $SU(N_c)$. Thus, it is obvious that
the product of two Fock space traces in (\ref{part2}) has to be a class 
function, which can be obtained below using the diagonalization condition
in maximal abelian sub-space of $SU(N_c)$.

Now, in each Fock space there exists a basis that diagonalizes 
both operators as long as ${\hat H}_i$ and ${\hat U}_i(g)$ commute. 
Let $|\alpha,\sigma\rangle$
be the one-particle states in such basis, where $\alpha$ labels the 
eigenvalues of ${\hat H}_i$ (including a possible degeneracy besides the one
associated with the symmetry group ${\cal G}$) and $\sigma$ labels those of 
${\hat U}_i(g)$. One can write the diagonalised eigenstate as
\begin{eqnarray}
\langle \alpha^\prime , \sigma^\prime | {\hat U}_i(g){\hat A}_i |\alpha , 
\sigma \rangle &=& \delta_{\alpha\alpha^\prime}
\delta_{\sigma\sigma^\prime}R_{i\sigma\sigma} A{_{i\alpha}}\ . \ \label{diag}
\end{eqnarray} 
where $A_{i\alpha}$ and $R_{i\sigma\sigma}(g)$ are, respectively, the 
eigenvalues of ${\hat A}_i$ and ${\hat U}_i(g)$. 
Then following  standard procedures one can obtain~\cite{aub,Gale} 
\begin{eqnarray}
{\rm{Tr}} \left ( {\hat U}_i e^{-\beta {\hat A}_i} \right ) {\hspace*{-0.1in}}
&=&{\hspace*{-0.1in}}
\exp{\left [ {\rm{tr}_{\alpha}\rm{tr}_c}  \ln \left (1+R_i(g)e^{-\beta A_{i\alpha }}
\right )\right ]} . \label{trace0}
\end{eqnarray}
Note that $`\rm{tr}_{\alpha}$' is the trace over the momentum state $\alpha$
whereas $`{\rm{tr}_c}$' is the trace over 
the colour degrees of freedom in the same momentum state $\alpha$. 
In fundamental representation 
${\rm{tr}_c} R(g^k)=\sum_{l=1}^{N_c}\exp (ik\theta_l)$ along with 
$R^k(g)=R(g^k)$. This can be related to the Polyakov 
Loop~\cite{pisar,polyakov} in Polyakov gauge as 
${\cal L}={\rm {tr}_c}(L)/N_c = {\rm{tr}_c} R(g^k)$, where $L$ is the thermal
Wilson lines defined by the temporal gluons in Euclidian time. 
This correspondence is due to the choice of diagonal $U(g)$, which resembles
the Polyakov Loop matrix in Polyakov gauge, supplemented
with the diagonalization condition in (\ref{diag}).

Now one can write the product of two traces in (\ref{part2}) as  
\begin{equation}
 {\rm{Tr}} \left ( {\hat U}_q(g) 
 e^{-\beta {\hat H}_q} \right )   
 {\rm{Tr}} \left ( {\hat U}_{\bar q}(g) 
 e^{-\beta {\hat H}_{\bar q}} \right ) 
= \ \ \exp \left ({\Theta }\right ) \ , \label{trace}
\end{equation}
where it is easy to show that
\begin{eqnarray}
\Theta &=& \sum_\alpha \sum^{N_c}_l\sum^{N_s}_s\sum^{N_f}_f \left [ 
\ln\left (1+e^{i\theta_l} e^{-\beta
(\epsilon_{q\alpha}-\mu_q)} \right ) \right.\nonumber \\ 
&& \ \ \ \ + \left. \ln\left (1+e^{-i\theta_l} e^{-\beta 
(\epsilon_{{ q}\alpha}+ \mu_{ q})} \right ) 
\right ] . \label{theta}
\end{eqnarray}
Here the flavour ($f$) and spin ($s$) summations 
are introduced, where  $N_f$ and $N_s$ are, respectively, the number of flavour 
and spin degrees of freedom. 
The equation (\ref{theta}) clearly indicates the 
superposition of two Fock spaces [fermionic and antifermionic determinants] 
with particles having same momentum and internal colour structure obeying the 
quantum statistics.  {\it The momentum states which do not satisfy this 
are automatically eliminated by the diagonalization condition in (\ref{diag})}. 
This essentially amounts to a stacking of same momentum particles.
It is also evident that even if we had started by considering a free gas of 
quarks and antiquarks, the colour-singlet restriction related to
the Polyakov Loop in Polyakov gauge links the effective degrees of freedom
(anti)quarks with the surrounding thermal bath through the temporal gluons.
As we will see, this allows an interesting scenario of recombinations. Now, 
the single-particle energy eigenvalues in a given state are same for $q$ 
and $\bar q$ as $\epsilon_{i\alpha}= \sqrt{p^2_{i\alpha}+m^2_i}$. 
However, their occupation energies in a given state differ by their chemical 
potentials as $\mu_{\bar q}=-\mu_q=-\mu$. For convenience~\cite{mus0}, we make 
a substitution $\xi = -i\beta \mu$ and $\epsilon_{q\alpha}=\epsilon_\alpha$ 
in (\ref{theta}) which then becomes  
\begin{eqnarray}
\Theta &=& \ln  \prod_\alpha \prod^{N_c}_l  \ {\cal D} \ \ , \nonumber \\
 \ {\cal D} &=& \Big [e^{-\beta \epsilon_{\alpha}}  
 \left ( 2 \cosh \beta \epsilon_{\alpha}
+2 \cos (\theta_l +\xi) \right ) \Big ]^{N_fN_s}, \label{simpl}
\end{eqnarray}
where the determinant ${\cal D}$ is a class function.
We would like to note that the class parameters in colour space get 
associated with the imaginary chemical potential $\xi$ due to the choice
of diagonal $U(g)$. As a consequence
of which we will see below that the colour factor $N_c$ appears with the
chemical potentials indicating the number of valence (anti)quarks. 

The partition function in (\ref{part2}) can now be written as
\begin{eqnarray}
{\cal Z}_S 
 &=&\prod_\alpha \int_{SU(N_c)} {\rm d}\mu (g) \ {\cal D} \ . 
\label{part3}
\end{eqnarray}
The integrations on class parameters in (\ref{part3})  
are now performed exactly by using the properties of the Jacobian 
and an orthonormal polynomial method~\cite{jacob}. The logarithm of the 
partition function~\cite{mus} for $N_c=3$ and two massless quarks ($N_f=2$) 
in the infinite volume, $V$, limit reads as 
\begin{eqnarray}
\frac{\ln{\cal Z}_S}{V} = \int \frac{{\rm d}^3p}{(2\pi)^3} \ln \left [1+
S \right ] \ \ , \label{colpart}
\end{eqnarray}
where $S$ is the sum of the Boltzmann factors of 
{\it the colour-singlet multiquark clusters} allowed by the symmetries
\begin{eqnarray}
S{\hspace*{-0.09in}}&=&\hspace*{-0.10in} \sum_{b=0}^{2N_f} 
\sum_{m=\delta_{0b}}^{6N_f-N_cb} 
{\hspace*{-0.1in}} C_{mb} 
\exp\left[{-(2m\epsilon +N_cb(\epsilon\mp\mu))\beta}\right ] 
. \label{colparts}
\end{eqnarray}
In doing so we have replaced  the $\sum_\alpha$ by the 
integration over phase space volume $d^3xd^3p/(2\pi)^3$. 
 It is interesting to note that due to the colour-singlet restriction the 
colour factor, $N_c=3$, and the baryon number, $b$, are always associated 
with the Boltzmann factor and thus with the partonic chemical potential 
indicating the excess number of quarks or antiquarks in addition to $m$ 
number of quarks and antiquarks. {\it This happens just because the diagonal 
matrix, $R_{\sigma \sigma}$, in colour space 
or the Polyakov Loop re-adjusts  the (anti)quarks of different momenta 
to be in a same momentum state 
by allowing them to exchange the momentum in
the thermal bath.}
As seen, in general, a given colour-singlet mode in (\ref{colparts}) has 
energy $E_{mb}=(2m +N_c b)\epsilon$ having parton content [$({m+N_cb})q, 
m{\bar q}$] for a hadron whereas that for anti-hadron is 
[$(m+N_cb){\bar q}, mq$].  The pure mesonic modes ($b=0$) can 
be identified with parton content ($mq, m{\bar q}$) having total energy 
$E_{m0}=2m \epsilon$. Similarly, the pure baryonic modes ($m=0$) have 
the energy $E_{0b}=N_cb\epsilon$ with parton content
[$N_cbq$] for baryon whereas that for antibaryons  is [$N_cb{\bar q}$].
In the above $C_{mb}$ is the weight factor, due to 
flavour and spin symmetry, appearing with each colour-singlet 
mesonic/baryonic/antibaryonic mode because of colour-singlet restriction.  
Their values for various modes are listed 
in Ref.~\cite{mus}. For low lying mesons ($m=1$ and $b=0$) $C_{10}=16$ 
whereas for low lying baryons and antibaryons ($m=0$ and $b=1$) $C_{01}=20$, 
respectively. These are exact for the  $SU(2)$ flavor and the $SU(2)$ spin 
symmetry~\cite{martin} of the
quark model in which nucleons and deltas are degenerate.
On the other hand
$m> 1$ and $b=0$ correspond to excited mesonic modes whereas $m\ge 1$ and $b
\ge 1$ are penta-quark and excited baryonic/antibaryonic modes, which are
the Hagedron states~\cite{hage}.

Equation (\ref{colpart}) clearly exhibits {\it a nontrivial} result: 
$SU(3)$ colour-singlet restriction, along with flavour and spin symmetry, on 
the quark-antiquark ensembles reorders the partition function in terms of 
Boltzmann factors of the colour-singlet multi-quark 
[mesonic/baryonic/antibaryonic] modes at any temperature. 
These are, however, not the bound states but can be regarded as a precursor 
to the confinement due to recombination~\cite{bass,greco,amati}. Under a 
suitable confining mechanism (e.g., Polyakov Loop model~\cite{pisar,polyakov}), 
one can hope that these multi-quark structures could 
evolve into colour-singlet hadrons in the low temperature limit.

\noindent{\it Probability and particle spectra:}
The probability of finding a single (anti)quark in the 
system in the energy interval $\epsilon$ and $\epsilon +d\epsilon$ 
follows from (\ref{colpart}) as~\cite{patha}
%\vspace{-0.20in}
\begin{eqnarray}
{\cal P}(\epsilon)d\epsilon &=& \frac{V d^3p}{(2\pi)^{3}} 
\ln \left [1+ S \right ] \ . 
\label{proba0}
\end{eqnarray}
The logarithm is expanded with $S<1$ that yields 
 a solution $\epsilon > \zeta T$, where $\zeta\sim 1.7$ for $N_f=2$. 
This provides an energy cut-off $\ge 2\zeta T$ for mesonic modes and 
$\ge 3\zeta T$ for baryonic modes.  One can write the above as
%\vspace*{-0.15in}
\begin{eqnarray}
{\cal P}(\epsilon)d\epsilon {\hspace*{-0.09in}}&=& {\hspace*{-0.1in}}
\sum_{b=0}^\infty \sum_{m=\delta_{0b}}^\infty {\hspace*{-0.1in}}
 C_{mb}^* \exp\left[{-(2m\epsilon +3b(\epsilon\mp\mu))\beta}\right] 
\frac{V d^3p}{(2\pi)^{3}} \nonumber \\
&=& \sum_{b=0}^\infty \sum_{m=\delta_{0b}}^\infty
 {\cal P}_{mb}^q(\epsilon) d\epsilon
\ , \label{proba}
\end{eqnarray}
in which all higher order terms are accumulated in $C_{mb}^*$, 
where $C_{mb}^*= C_{mb}$ for only low lying hadronic modes. 
In the above ${\cal P}_{mb}^q(\epsilon)d\epsilon$ is the probability of a 
single parton, with energy $\epsilon>\zeta T$, in the interval 
$\epsilon$ and $\epsilon + d\epsilon$ in a given $mb-$th mode which reads
as 
\begin{eqnarray}
{\cal P}_{mb}^q(\epsilon) d\epsilon = C_{mb}^* 
\exp\left[{-(2m\epsilon +3b(\epsilon\mp\mu))\beta}\right] 
\frac{V d^3p}{(2\pi)^{3}} . \label{probmb}
\end{eqnarray}
Now, the distribution of a parton within a given $mb-$th mode in the fluid in 
terms of the momentum of the parton ($p^\mu u_\mu=\epsilon$, $u_\mu$ is
4-velocity of the fluid) 
follows directly from (\ref{probmb}) as
\begin{equation}
%\frac{{\cal P}_{mb}^q(\epsilon )d\epsilon}{d^3x d^3p}\equiv
\frac{dN_{mb}^q}{d^3x \ d^3p} = \frac{C_{mb}^*}{(2\pi)^3}\ 
\exp\left [{-(2m\epsilon +3b(\epsilon\mp\mu))\beta}\right]
 \ \ . \label{distmb}
\end{equation} 
One can easily invert the parton momentum distribution in the above 
into the momentum [$P^\mu u_\mu=E_{mb}=(2m+3b)\epsilon$] distribution of 
a $mb-$th hadronic mode
as
\begin{eqnarray}
\frac{dN_{mb}}{d^3x d^3P} &=& 
n^3 \frac{dN_{mb}^q}{d^3x d^3P} = 
\frac{C_{mb}^*}{(2\pi)^3} 
 e^{-(E_{mb}\mp 3b\mu)\beta}, \label{distmbh}
\end{eqnarray} 
where $n=(2m+3b)$ and the above is the Cooper-Frye distribution~\cite{coop}  
at the freeze-out in the rest frame of the  fluid.
We assume that this  distribution describes the corresponding hadrons 
after freeze-out.
The above distribution has also an important consequence: the entropy would
remain conserve since the number of quarks in quark's phase space is equal
to that of hadrons in hadron's phase space. 

Now we can obtain the differential proton to pion ratio at central rapidity
for a given transverse momentum $P_\bot >> 3\zeta T$ as
\begin{eqnarray}
\frac{dN_{01}}{dN_{10}}= \frac{dN_{p}}{dN_{\pi}}= \frac{C^*_{01}}{C^*_{10}}
e^{3\mu/T}= \frac{5}{4}e^{\mu_B/T} \ , \, \label{ppir}
\end{eqnarray}
where $\mu_B=3\mu$, is the baryonic chemical potential and the above is
independent of $P_\bot$.  
PHENIX data show~\cite{phenix} that the ratio has a plateau around 
unity in the $P_\bot$ range, 
$ 2 {\ \rm{GeV}}/c \leq P_\bot \leq 4 \ {\rm{GeV}}/c$.
Our estimation shows that it is in good agreement with RHIC 
data~\cite{phenix,star}. 
Recall that in the recombination model by Fries et al.~\cite{bass} 
it was found to be $(5/3)\exp(\mu_B/T)$. Similarly, antiproton to pion 
is $(5/4)\exp(-\mu_B/T)$ and that of antiproton to proton is 
$\exp({-2\mu_B/T})$, which are again in good agreement with the RHIC 
data~\cite{adler}. Thus, the colour-singlet projection of the thermal parton 
ensemble based on symmetry consideration of the underlying theory also 
provides a strong basis for the hadronization by recombination
of quark-gluon plasma at intermediate $P_\bot$ 
($ 2 {\ \rm{GeV}}/c \leq P_\bot \leq 4 \ {\rm{GeV}}/c$) even though the
ingredients are different from the recombination models~\cite{bass,greco} 
with a sudden approximation.
 
Now the total number of a given hadron  emitted by a fluid can be obtained
from (\ref{distmb}) as
\begin{eqnarray}
N_{mb}&=& \frac{VC_{mb}^*}{(2\pi)^3} T^3 
e^{-n\zeta \pm 3b\mu/T} 
\left[ 1+ n \zeta + \frac{n^2\zeta^2}{2}\right ] , 
\label{hadnum}
\end{eqnarray}
which simply depends on the number of partonic degrees of freedom that forms
the colour-singlet hadronic modes with $\epsilon/T > \zeta$. 
The ratio of the proton to pion becomes
$\sim 0.6 $ whereas it is $\sim 0.3$ for antiproton to pion with 
$\mu_B/T\sim 0.33$ in RHIC. This is consistent with the RHIC 
data~\cite{phenix}. 

\noindent{\it Anisotropy and scaling:} 
We now  consider non-central collisions where the fluid  
has larger velocity on the x-axis (semi-minor) than on the 
y-axis (semi-major) leading to elliptic flow~\cite{olli}. 
The flow coefficient is defined as
%\vspace*{-0.08in}
\begin{equation}
v_2 = \langle \cos2\phi \rangle \, \ , \label{coeff}
\end{equation} 
This requires that the distribution in (\ref{distmb}) should have a nontrivial
dependence on azimuthal angle $\phi$. It is introduced through  
$\epsilon=p^\mu u_\mu=p_\bot u_0(\phi) -p_\bot u(\phi)$, where $u_\mu$ is 
the 4-velocity of the fluid and  $u(\phi)$ can be 
parameterized~\cite{olli} in the following form
\begin{equation}
u(\phi)= u+2\alpha \cos2\phi \ \ , \ \ \label{uphi}
\end{equation}
where $u$ is $\phi-$averaged of the maximum fluid $4-$velocity in the $\phi$
direction and $\alpha$ 
specifies the magnitude of the elliptic flow, which is about 
$4\%$ in a non-central Au-Au collisions at RHIC.
Considering $u^\mu u_\mu = u_0^2(\phi)-u^2(\phi)= 1$ and expanding it to
a first order in $\alpha$, one can obtain
\begin{equation}
u_0(\phi)=u_0+2\alpha  v  \cos 2\phi \ \ , \label{u0phi}
\end{equation}
where $v\equiv u/u_0$.  Now, the elliptic flow coefficient for the 
$mb-$th mode in (\ref{coeff}) is found to scale perfectly with 
the partonic $p_\bot$ as
\begin{equation}
(v_2)_{mb} = \frac{\alpha}{T}\left(1-v \right ) \left(2m+3b\right ) 
p_\bot \ .  \label{scale}
\end{equation} 

We note that this scaling is strictly valid in the recombination region
($ 2 {\ \rm{GeV}}/c \leq P_\bot \leq 4 \ {\rm{GeV}}/c$), where the mass 
dependence of the hadrons is irrelevant. However, for 
$P_\bot < 1.5$GeV/$c$ the scaling deviates~\cite{ellip} due to the mass
ordering of the hadrons. To demonstrate this fact one needs a dynamical
mass generating term of the quarks, which is, however, beyond scope 
of this calculation.

\noindent{\it Summary:}
We propose that the recombination
phenomena occurs readily when one constructs a colour-singlet partition
function from a thermal ensemble of quarks and antiquarks having internal
symmetries, {\it viz.}, colour, spin and flavour. This colour-singlet 
projection is shown to be related to the Polyakov Loop in the Polyakov gauge, 
which plays the important role for recombination. We also show that such
recombination of thermal partons naturally describes some of the puzzling 
results in RHIC experiments, which, in turn, strongly supports that the 
recombination of thermal (anti)quarks could be one of the dominant mechanisms 
of hadronization in a dense medium in the intermediate transverse momentum 
of hadrons ($ 2 {\ \rm{GeV}}/c \leq P_\bot \leq 4 \ {\rm{GeV}}/c$). 
Nevertheless, to test the various quark recombination models for hadronisation 
from dynamical QCD still remain an open question. We, however, tried to 
address this question based on the symmetry consideration of the theory, which 
may be useful for the eventual solution of this important problem. 

\noindent{\it Acknowledgment:} Valuable
discussions with B. M\"uller,  D. K. Srivastava, S. Bass and R. Ray are
acknowledged.


\begin{thebibliography}{99}
\bibitem{white} BRAHMS Collaboration, Nucl. Phys. A {\bf 757}, 1 (2005); 
PHENIX Collaboration, {\it ibid} {\bf 757}, 184 (2005); 
PHOBOS Collaboration, {\it ibid} {\bf 757}, 28 (2005);
STAR Collaboration, {\it ibid} {\bf 757}, 102 (2005).

\bibitem{ellip} PHENIX Collaboration A. Adare {\it et al.},  
Phys. Rev. Lett.{\bf 98}, 162301 (2007)

\bibitem{jet} PHENIX Collaboration, K. Adcox {\it et al.}, Phys. Rev. 
Lett.{\bf 88}, 022301 (2002); STAR Collaboration, C. Adler {\it et al.},
Phys. Rev. Lett. {\bf 89}, 092302 (2002).

\bibitem{phot} PHENIX Collaboration, S. S. Adler {\it et al.},
Phys. Rev. Lett. {\bf 98}, 012002 (2007).

\bibitem{phenix} PHENIX Collaboration, S. S. Adler {\it et al.}, 
Phys. Rev. Lett. {\bf 91}, 172301 (2003); T. Chujo, PHENIX Collaboration, 
Nucl. Phys. A{\bf 715}, 151c (2003). 

\bibitem{bass} R. J. Fries {\it et al.},
Phys. Rev. Lett. {\bf 90}, 202303 (2003); Phys. Rev. C{\bf 68}, 044902 (2003).

\bibitem{greco} V. Greco, C. M. Ko, and P. Levai, Phys. Rev. Lett. {\bf 90}, 
202302 (2003).

\bibitem{rec} T. S. Bir\'o, P. L{\'e}vai and J. Zim\'anyi, Phys. Lett. 
B{\bf 347}, 6 (1995); Phys. Rev. C{\bf 59}, 1574 (1999); T. S. Bir\'o, 
T. Cs\"org\"o, P. L{\'e}vai and J. Zim\'anyi, Phys. Lett. B{\bf 472}, 243 
(2002); P. Csizmadia and P. L{\'e}vai, Phys. Rev. C{\bf 61}, 031903 (2000) 

\bibitem{rec1} B. Zhang {\it et al.}, Phys. Rev. C{\bf 61}, 067901 (2000);
Z. Lin {\it et al.}, Phys. Rev. C{\bf 64}, 011902 (2001).

\bibitem{cmko} Z. Lin and C. M. Ko, Phys. Rev. C{\bf 65}, 034904 (2002);
Phys. Rev. Lett. {\bf 89}, 152301 (2002); Phys. Rev. Lett. {\bf 89}, 202302
(2002).

\bibitem{weyl} H. Weyl, {\it The Classical Groups} (Princeton U.P., 
Princeton, NJ, 1946); F. D. Murnghan, {\it The Theory of Group
Representation} (Johns Hopkins University, MD, 1938).

\bibitem{aub} G. Auberson {\it et  al.}, J. Math. Phys. {\bf 27}, 1658
(1986).

\bibitem{Gale} J. I. Kapusta and C. Gale, {\it Finite-Temperature
Field Theory Principles and Applications}, Second Edition 
(Cambridge University Press, Cambridge, England, 1996)

\bibitem{red} I. Zakout and C. Greiner, Phys. Rev. C{\bf 78}, 034916 (2008);
B. M\"uller, {\it The Physics of Quark-Gluon Plasma}, First
Edition (Springer-Verlag, Berlin, Germany, 1985); 
K. Redlich and L. Turko, Z. Phys. C{\bf 5}, 201 (1980);
L. Turko, Phys. Lett. B{\bf 104}, 153 (1981); H-T Elze and W. Greiner, 
Phys. Lett. B{\bf 179}, 385 (1986); M. I. Gorenstein {\it et al.}, 
Phys. Lett. B{\bf 123}, 437 (1983); B. Skagerstam, Z. Phys. C{\bf 24}, 
97 (1984); K. Kusaka, Phys. Lett. B{\bf 269}, 17 (1991); A. Ansari and 
M. G. Mustafa; Nucl. Phys. A{\bf 539}, 752 (1992); 
M. G. Mustafa {\it et al.}, Phys. Lett. B{\bf 311}, 
277 (1993); M. G. Mustafa, Phys. Lett.  B{\bf 318}, 517 (1994);
M.G. Mustafa and A. Ansari, Phys. Rev. C{\bf 55}, 2005 (1997); 
M. G. Mustafa, D. K. Srivastava, and B. Sinha, Eur.Phys. J. C{\bf 5}, 711
(1998). 

\bibitem{mus0} M. G. Mustafa, Phys. Rev. D{\bf 49}, 4634 (1994).

\bibitem{mus} M.G. Mustafa, A. Sen, and L. Paria, Eur.Phys. J. C{\bf 11}
729, (1999).

\bibitem{jacob} P. H. Damgaard, N. Kawamoto, and K. Shigemoto, Nucl. Phys.
B{\bf 264}, 1 (1986); Y. Y. Goldschmidt, J. Math. Phys. {\bf 21}, 1842 (1980).

\bibitem{pisar} R. D. Pisarski, Phys. Rev. D{\bf 62}, 111501(R) (2000).

\bibitem{polyakov} K. Fukushima, Phys. Lett. B{\bf 591}, 277 (2004); Phys. Rev.
D{\bf 77}, 114028 (2008); C. Ratti, M. A. Thaler and W. Weise, Phys. Rev. 
D{\bf 73}, 014019 (2006); S. K. Ghosh {\it et al.} Phys. Rev. 
D{\bf 73}, 114007 (2006); {\it ibid} {\bf 77}, 094024 (2008).

\bibitem{martin} D. B. Lichtenberg, {\it Unitary Symmetry and Elementary
Particles}, Second Edition (Academic Press, INC. (London) Ltd.  1978).

\bibitem{hage} R. Hagedron, Nuovo Cimento Suppl. {\bf 3} (1965) 147.

\bibitem{amati} D. Amati and G. Veneziano, Phys. Lett. B{\bf 83}, 87 (1979).


\bibitem{patha} R. K. Patharia, {\it Statistical mechanics},
First Edition (Pergamon, Oxford, England, 1972)

\bibitem{coop} F. Cooper and G. Frye, Phys. Rev. D{\bf 10}, 186 (1974).

\bibitem{star} G. J. Kunde, STAR Collaboration, Nucl. Phys. A{\bf 715}, 189c 
(2003). 

\bibitem{adler} STAR Collaboration, C. Adler {\it et al.}, Phys. Rev. Lett.
{\bf 86}, 4778 (2001).

\bibitem{olli} J. Y. Ollitrault; Phys. Lett. B{\bf 273}, 32 (1991); 
Phys. Rev. D{\bf 46}, 229 (1992);  arXive:0708.2433 [hep-ph].
\end{thebibliography}
\end{document}